\documentclass{physeauth}
\usepackage{graphicx}
\usepackage{amsmath}
\usepackage{amssymb}

 \def\bra#1{\langle #1|}
 \def\ket#1{|#1\rangle}
 \def\sbra#1{\langle\kern-.2em\langle #1|}
 \def\sket#1{|#1\rangle\kern-.2em\rangle}
 
 \def\sbraket#1#2{\langle\kern-.2em\langle#1|#2\rangle\kern-.2em\rangle}
 \def\super#1{\mathcal{#1}}
 \def\L{{\rm L}}
 \def\R{{\rm R}}

\def\obrazek[#1(#2),#3.eps,#4]#5{
 \begin{figure}[tb]
   \begin{center}
        \includegraphics[angle=0,width=.47\textwidth]{#3.eps}
     \caption{#5}
     \label{#4}
   \end{center}
 \end{figure}
 }

 \let\epsilon\varepsilon
 \def\_#1{_{\rm #1}}         
 \def\tr{\mathop{\rm Tr}\nolimits\!}   
 \def\im{{\rm i}}            
 \def\eu{{\rm e}}            
 \def\d{\mathop{{\rm d}\!}
     \vphantom{\rm d}}
 \def\der#1#2{\frac{\d #1}{\d #2}}   
 
 \def\({\left(} \def\){\right)}      
 \def\cotgh{\mathop{\rm coth}\nolimits}      

 \def\E #1 {\cdot 10^{#1}}                       
 \bgroup\uccode`~=`e\uppercase{\egroup\let~}\E   
 \bgroup\uccode`~=`"\uppercase{\egroup\def~}#1"{\jedA #1 @@}
 \mathcode`\"="8000      
 \def\jedA #1 {\bgroup\mathcode`\,="013B \bgroup \mathcode`\e="8000 #1 \jedB}
 \def\jedB #1@@{\egroup\if @#1@\else\,\rm\mathcode`\.="0201 #1\fi\egroup}
\begin{document}

 \renewcommand{\arraystretch}{1}

\begin{frontmatter}

\title{Charge conservation breaking within generalized master equation description of electronic transport through dissipative double quantum dots}

\author[uk]{Jan Pracha\v r} and
\author[uk]{Tom\'a\v s Novotn\'y \thanksref{thank1}}

\address[uk]{Department of Condensed Matter Physics, Faculty of Mathematics and
Physics, Charles University in Prague, Ke~Karlovu~5, 121~16 Praha~2,
Czech Republic}

\thanks[thank1]{
Corresponding author. E-mail: tno@karlov.mff.cuni.cz}

\begin{abstract}
We report an observation of charge conservation breaking in a model
study of electronic current noise of transport through a dissipative
double quantum dot within generalized master equation formalism. We
study the current noise through a~double quantum dot coupled to two
electronic leads in the high bias limit and a~dissipative heat bath
in the weak coupling limit. Our calculations are based on the
solution of a~Markovian generalized master equation. Zero-frequency
component of the current noise calculated within the system, i.e.,
between the two dots, via the quantum regression theorem exhibits
unphysical negative values. On the other hand, current noise
calculated for currents between the dots and the leads by the
counting variable approach shows no anomalies and seems physically
plausible. We inquire into the origin of this discrepancy between
two nominally equivalent approaches and show that it stems from the
simultaneous presence of the two types of baths, i.e., the
electronic leads and the dissipative bosonic bath. This finding
raises interesting questions concerning conceptual foundations of
the theory describing multiple-baths open quantum systems widely
encountered in nanoscience.
\end{abstract}

\end{frontmatter}


\section{Introduction}

Recent advances in technology, fabrication, and measurement of
mesoscopic semiconductor devices with ever-decreasing dimensions of
achievable nanostructures stimulate also theoretical studies of
physical phenomena determining their properties. Questions of prime
interest concern their possible quantum behavior and the
quantum-classical crossover due to interaction with surrounding
environment causing dissipation, relaxation, and dephasing
\cite{weiss,gardiner,ankerhold}. One of the conceptually simplest
and experimentally achieved systems is the double quantum dot (DQD)
which can be tuned into a regime where it is effectively described
as a~tunable two-level system for the electronic energy
states~\cite{exp}. These energy states can be tuned, for instance,
by means of an external gate voltage. The interest in such devices
stems also from the attractive possibility to utilize them as
potential q-bits.

In the setup consisting of a DQD, the role of the electronic
coherence between the two spatially separated electronic states
corresponding to the respective dots is of central importance. The
DQD device loses coherence due to the coupling to noisy environment
(e.g.\ noise in the gate voltages and unavoidable interaction with
phonons in the substrate). Moreover, energy can be exchanged with
bosonic degrees of freedom which gives rise to transitions between
states of nonequal energy and, thus, relaxation. The dissipative
dynamics of two level systems (spin-boson problem) have been subject
of study for many years~\cite{weiss,leggett}. DQD setup brings along
a new twist to this standard problem in the fact that the DQD is
electrically contacted by leads and charge transport through the DQD
occurs \cite{exp}. This is a new feature of the old problem adding
complexity to the methods of its solution. On the other hand, for
most practically interesting setups the dissipative coupling may be
considered as rather weak which should bring about important
simplifications for its solution. Yet, as we will demonstrate in
this work, the simultaneous presence of the two different kinds of
baths (electronic leads and standard dissipative bosonic bath \`{a}
la Caldeira-Leggett \cite{caldeira}) causes serious conceptual
problems within the simplest possible formalism of Markovian
generalized master equations (GME) when it is applied to the weakly
dissipative DQD problem.

It has been suggested by Aguado and Brandes \cite{aguado} that
considering the electronic current noise in the DQD devices may be a
useful tool for characterizing their dissipative properties going
beyond the information available from the stationary current
characteristics only. In this work we adopt their model and perform
an exhaustive comparative study of the evaluation of the current
noise based on GME approach used in previous
studies~\cite{aguado,novotny}. We surprisingly find an internal
inconsistency of the formalism which breaches fundamental physical
law of charge conservation. Two nominally equivalent approaches for
the calculation of the zero-frequency component of the current noise
spectrum, namely the quantum regression theorem (QRT) and counting
variable approach with the MacDonald formula, show mutual
discrepancy. This fact puts under question the status of the results
obtained by these methods.

The structure of the paper is the following. In Sec.~\ref{model} we
introduce our model and in Sec.~\ref{gme} we describe the method of
its solution via an approximate Markovian GME. The following
Sec.~\ref{current_noise} describes the evaluation of the current
noise within the used GME formalism while
Sec.~\ref{charge_conservation} discusses the general aspects of the
charge conservation and its breaking within the GME formalism.
Sec.~\ref{results} presents an overview of results for our studied
model of a dissipative DQD. We discuss the obtained results and
their implications together with resulting open problems and outlook
in the concluding Sec.~\ref{conclusions}.

\section{Model}
\label{model}

The double quantum dot device~\cite{exp} in Fig.~\ref{fig1_system}
is described as two electronic levels (corresponding to particular
single-electron levels within the transport window of the left and
right dot, respectively) de-aligned by an~energy
difference~$\epsilon$ with a~coherent interdot tunnel
coupling~$\Omega$. The system is assumed to be in the regime of
strong Coulomb blockade so that only three states play a role: no
extra electron~$\ket0$ on the whole DQD system, one extra electron
on the left dot~$\ket \L$ and one extra electron on the right
dot~$\ket \R$, i.e., we exclude multiple occupancies of the DQD.
This can be achieved by a~suitable gating, when a very high charging
energy prohibits an addition of more than one electron.  We also
consider spinless electrons. Hamiltonian of the DQD device then
reads
\begin{equation}
    H\_S = \tfrac12\epsilon\big(\ket\L\bra\L-\ket\R\bra\R\big)+ \Omega\big(\ket \L\bra \R + \ket \R\bra \L\big).
    \label{hamiltonian_S}
\end{equation}
The so called device bias~$\epsilon$ can be tuned by gating. The
term proportional to~$\Omega$ enables the tunnel current through the
device. The eigenvalues of the isolated system Hamiltonian are
$E_{1,2}=\pm\frac12\Delta$ with $\Delta =
\sqrt{4\Omega^2+\epsilon^2}$ and the corresponding eigenvectors read
$\ket
1=\sqrt{\tfrac{\Delta+\epsilon}{2\Delta}}\ket\L+\sqrt{\tfrac{\Delta-\epsilon}{2\Delta}}\ket\R,\,
\ket
2=-\sqrt{\tfrac{\Delta-\epsilon}{2\Delta}}\ket\L+\sqrt{\tfrac{\Delta+\epsilon}{2\Delta}}\ket\R$.

\obrazek[MP(),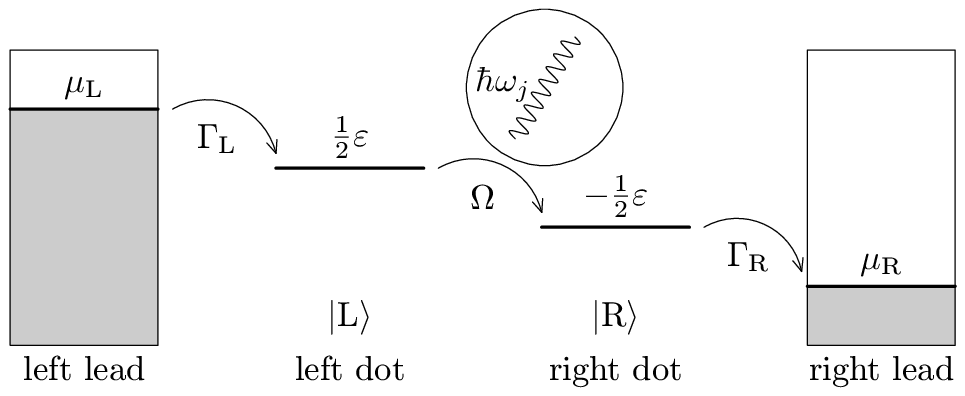,fig1_system]{Schematic depiction of the
studied DQD system.}

The double quantum dot is coupled to two leads with a~high bias
applied between them. The bias is smaller than the charging energy
but otherwise it is the largest energy scale in the model. We assume
that the noninteracting leads are coupled via standard tunneling
terms
\begin{equation}
\begin{aligned}
    H\_C+H\_{CS} &= \sum_{k}E_{k\L} c_{k\L}^\dagger c_{k\L}+\sum_{k}E_{k\R} c_{k\R}^\dagger
    c_{k\R}\\
    &+\sum_{k}V_{k\L}\big(c_{k\L}^\dagger \ket 0\bra\L + \ket\L\bra0 c_{k\L}\big)\\
    &+\sum_{k}V_{k\R}\big(c_{k\R}^\dagger \ket 0\bra\R + \ket\R\bra0 c_{k\R}\big).
\end{aligned}
\label{hamiltonian_CCS}
\end{equation}
The leads are held at respective electrochemical potentials $\mu_\L$
and $\mu_\R$ whose difference gives the bias. We assume that
$\mu_\L\to\infty$ and $\mu_\R\to-\infty$. The tunneling densities of
states $\Gamma_\alpha(\epsilon) =
\frac{2\pi}\hbar\sum_k|V_{k\alpha}|^2\delta(\epsilon-E_{k\alpha})\
,\alpha=\L,\R$ are assumed energy independent (wide-band limit or
first Markov approximation, compare Ref.~\cite{gardiner}) and equal
$\Gamma_\L = \Gamma_\R=\Gamma$. Both the high bias and wide-band
limits are necessary for the applicability of the Markov
approximation later on.

Finally, we introduce a generic dissipative heat bath \`{a} la
Caldeira-Leggett consisting of an infinite set of harmonic
oscillators \cite{weiss,caldeira} which are linearly coupled to the
left-right  population difference of the DQD \cite{aguado}
\begin{equation}
    \begin{aligned}
    H\_B+H\_{BS} &= \sum_j\hbar\omega_j(a_j^\dagger
    a_j+\tfrac12)\\
    &+
    \sum_j C_j(a_j^\dagger+a_j)(\ket\L\bra\L-\ket\R\bra\R).
    \end{aligned}
    \label{hamiltonian_BBS}
\end{equation}
The heat bath is fully characterized by its spectral density
\begin{equation}
    J(\omega) = 2\sum_j|C_j|^2\delta(\omega-\omega_j),
    \label{bath_spectral_density}
\end{equation}
which we take in the Ohmic form $J(\omega) =
2\hbar^2\gamma\omega/\pi\cdot\exp{(-\omega/\omega\_c)}$. The
parameter~$\gamma$ gives the strength of the dissipation and
$\omega\_c$ is a~high energy cut-off frequency
\cite{weiss,caldeira}.

\section{Generalized master equation}
\label{gme}
\subsection{Liouville space}

In order to conveniently manipulate with density operators, we
define Liouville space \cite{novotny}. The Liouville space is
a~linear space spanned over operators acting on the original Hilbert
space assigned to the system. Its basis~$\sket{n,n'}$ is constructed
from a~basis~$\ket{n}$ of the Hilbert space as $\sket{n,n'} \equiv
\ket n\bra{n'}$. A general operator in the Hilbert space
$$
    A= \sum_{n,n'}A_{nn'}\ket n\bra{n'}
$$
corresponds to the vector
$$
    \sket A  = \sum_{n,n'}A_{nn'}\sket{n,n'}
$$
in the Liouville space. For example, a density operator $\rho$ can
be written as $\rho = \sum_{n,n'}\rho_{nn'}\sket{n,n'}$ with
$\rho_{nn'}=\bra n \rho\ket{n'}$. We define the scalar product on
the Liouville space by $\sbraket{A}{B} \equiv \tr{\{A^\dagger B\}}$.
In order to avoid confusion, linear operators acting in the
Liouville space are called superoperators and in the following will
be denoted by calligraphic symbols. The vectors of the Liouville
space in the bra-ket notation will be distinguished by double
brackets. The matrix representation of superoperators then follows
from the previous
$$
    \super A =
    \sum_{\substack{n,n'\\ m,m'}}A_{nn',mm'}\sket{n,n'}\sbra{m,m'}.
$$

\subsection{Liouvillean}

For the description of a dissipative system we distinguish between
the system itself (electronic states of the dots) and the reservoirs
(heat bath and leads). Our task now is to get a closed evolution
equation for the reduced density operator which is the system part
only of the total density operator. To this end we perform the
standard projection onto the system assuming weak coupling to the
reservoirs and consequently using the Markov approximation
\cite{gardiner}. Due to the weak coupling the effects of the two
baths are additive.

It should be noted that within our assumptions on the leads (wide
band limit and high bias) an equivalent result for the effect of
leads can be obtained without the weak coupling assumption, i.e.,
for arbitrary $\Gamma$, as shown by Gurvitz and Prager
\cite{gurvitz}. It turns out that these assumptions correspond
exactly to the so called singular coupling limit in the mathematical
literature (see, e.g., Ref.~\cite{spohn}) which also leads to a
Markovian dissipative evolution of the system. The additivity of the
two bath is then, however, only a heuristic assumption which may
break down for large enough $\Gamma$. The range of parameters we
consider is restricted to rather small $\Gamma$ so that the
potential differences between the two possible approaches should be
safely negligible and the result of the projection leads to the
following GME
\begin{equation}
    \der{\rho(t)}{t} = \super L\rho(t) =\super L\_S\rho(t)+\super L\_B\rho(t)+\super L\_C\rho(t),
    \label{dqd:gme}
\end{equation}
with
\begin{align*}
    \super L\_S\rho(t) &= -\frac\im\hbar[H\_{S},\rho(t)],\\
    \super L\_B\rho(t) &= -\frac1{\hbar^2}\int_0^\infty\d\tau\\
    &\hskip4em\times\tr\_B\{[H\_{BS},[H\_{BS}(-\tau), \rho(t)\otimes\rho\_B]]\},\\
    \super L\_C\rho(t) &= -\frac1{\hbar^2}\int_0^\infty\d\tau\\
    &\hskip4em\times\tr\_C\{[H\_{CS},[H\_{CS}(-\tau), \rho(t)\otimes\rho\_C]]\}.
\end{align*}

The part~$\super L\_S$ describes the free evolution of the system
while $\super L\_B,\,\super L\_C$ determine the dissipative
influence of the generic heat bath and the electronic leads,
respectively. It turns out that the off-diagonal elements
$\rho_{0k}$, $\rho_{k0}$ with $k =1,2$  of the reduced density
matrix are decoupled from the rest of the system, i.e., their
evolution does not enter expressions for the other matrix elements
and vice versa (compare with Ref.~\cite{novotny}). Therefore, the
subspace $\{\sket{0\alpha},\sket{\alpha0}\}$ ($\alpha = \L,\R$) can
be projected out leaving us with the relevant Liouville subspace
with the basis
$\{\sket{00},\sket{\L\L},\sket{\R\R},\sket{\R\L},\sket{\L\R}\}$. In
this basis the above parts of the total Liouvillean are described by
the following matrices \cite{aguado,brandes}
\begin{align}
    \super L\_S &= \frac1\hbar
    \begin{pmatrix}
    0 & 0 & 0 & 0 & 0 \\
    0 & 0 & 0 & -\im \Omega & \im \Omega\\
    0 & 0 & 0 & \im \Omega & -\im \Omega \\
    0 & -\im \Omega & \im \Omega & \im\epsilon & 0  \\
    0 & \im \Omega & -\im \Omega & 0 & -\im\epsilon
    \end{pmatrix},
    \label{liouvillian_S_LR}\\
    \super L\_B &= \begin{pmatrix}
    0 & 0 & 0 & 0 & 0 \\
    0 & 0 & 0 & 0 & 0 \\
    0 & 0 & 0 & 0 & 0 \\
    0 & \gamma_+& -\gamma_-& -\gamma_p& 0 \\
    0 & \gamma_+& -\gamma_-& 0 & -\gamma_p\\
    \end{pmatrix},
    \label{LiouvillianB_LR}\\
    \super L\_C &= \begin{pmatrix}
    -\Gamma & 0 & \Gamma & 0 & 0 \\
    \Gamma & 0 & 0 & 0 & 0 \\
    0 & 0 & -\Gamma & 0 & 0 \\
    0 & 0 & 0 & -\tfrac12\Gamma & 0 \\
    0 & 0 & 0 & 0 & -\tfrac12\Gamma\\
    \end{pmatrix},
    \label{LiouvillianL_LR}\\
    \intertext{with}
    \gamma_{\pm} &= -\frac\pi{\hbar^2}\frac{\Omega}{\Delta}\,J(\Delta/\hbar)
    \left[\frac{\epsilon}{\Delta}\cotgh{(\tfrac12\beta\Delta)}\pm1\right],\\
    \gamma_{p} &=
    \frac{4\pi}{\hbar^2}\frac{\Omega^2}{\Delta^2}J(\Delta/\hbar)\cotgh{(\tfrac12\beta\Delta)}.
\end{align}

\section{Current noise}
\label{current_noise}

In this section we briefly introduce the current noise
\cite{blanter,robinson} and methods of its evaluation within the GME
framework. More detailed account of these issues can be found in
Sec. III of Ref.~\cite{novotny}.

Equations of motion for the operators of the occupation of the left
dot $n_\L = \ket\L\bra\L$ and the right dot $n_\R = \ket\R\bra\R$
read
\begin{align}
    e\,\der{}tn_\L &= -\frac{\im e}{\hbar}[n_\L, H] =
    I_{\L0}-I_{\R\L},
    \label{eom1_dqd}\\
    e\,\der{}tn_\R &= -\frac{\im e}{\hbar}[n_\R, H] = I_{\R\L}-I_{0\R}.
    \label{eom2_dqd}
\end{align}
On the right side of the equations, we identify charge current
operators across the different junctions: $I_{\L0}=-\tfrac{\im
e}{\hbar}[n_\L, H\_{CS}]$ is the operator of the current between the
left lead and the left dot, $I_{\R\L}= \tfrac{\im e}{\hbar}[n_\L,
H\_{S}]=-\tfrac{\im e}{\hbar}[n_\R, H\_S]$ is the operator of the
current between the dots, and $I_{0\R}= \tfrac{\im e}{\hbar}[n_\R,
H\_{CS}]$ is the operator of the current between the right dot and
the right lead. Explicitly, they read
\begin{align}
    I_{\L0} &=  \frac{\im e}{\hbar}\sum_kV_{k\L}\big(c_{k\L}^\dagger\ket0\bra\L-\ket\L\bra0
    c_{k\L}\big),\\
    I_{\R\L}&= \frac{\im e}{\hbar}\Omega\big(\ket\L\bra\R-\ket\R\bra\L\big),\\
    I_{0\R} &= \frac{\im e}{\hbar}\sum_kV_{k\R}\big(\ket\R\bra0 c_{k\R}-c_{k\R}^\dagger\ket0\bra\R\big).
\end{align}
Since the commutators with the bath operators are zero $[n_\L,
H\_{BS}]=[n_\R, H\_{BS}] =0$, the heat bath gives no explicit
contribution to the current operators.

The current operator~$I_{\R\L}$ is obviously a system operator,
i.e., it acts as unity on the degrees of freedom of the leads and
the heat bath. However, this is not the case for the two operators
of current between the dots and the leads $I_{\L0}$ and~$I_{0\R}$.

Next, we define the current autocorrelation function
\begin{equation}
    \begin{aligned}
    C_{A}(\tau) &\equiv \lim_{t\to\infty}\left[\tfrac12\langle \{I_A(t+\tau),I_A(t)\}\rangle\right.\\
    &\hskip3em\left.\vphantom{\tfrac12}-\langle I_A(t+\tau)\rangle\langle
    I_A(t)\rangle\right],
    \end{aligned}
    \label{definition_C_A}
\end{equation}
with $A =\L0,\,\R\L,\,0\R$. Due to the stationary limit
($t\to\infty$) the autocorrelation function is symmetric $C_A(\tau)
= C_A(-\tau)$. We define the current noise spectrum as
\begin{equation}
    S_A(\omega) \equiv \int_{-\infty}^\infty\d\tau C_A(\tau)\eu^{\im\omega\tau}\,.
    \label{definition_noise}
\end{equation}
The current noise spectrum is non-negative as can be shown by using
the Lehmann representation.

Now we need to express the current noise spectrum in terms of the
quantities involved in the GME \eqref{dqd:gme}. We denote the
stationary reduced density matrix
$\lim_{t\to\infty}\rho(t)=\rho\_{stat}\equiv \sket0$. It satisfies
$\super L\rho\_{stat}=0$, hence it is the zero-eigenvalue (right)
eigenstate of the Liouvillean. Since the Liouvillean is not
Hermitian, left zero-eigenvalue eigenstate denoted by $\sbra{\tilde
0}$ is not just the Hermitian conjugate of the right zero-eigenvalue
eigenstate $\sket0$. However, one can see that $\sbra{\tilde0}\equiv
1$, because for an arbitrary system operator~$A$
$$
    \sbra{\tilde 0}\super L\sket A = \tr\_S\big(1\super LA\big)= \tr\_S\big(\super L
    A\big)=0
$$
due to normalization of the reduced density matrix. Now, we define
the projector on the kernel of the Liouvillean $\super P\equiv
\sket0\sbra{\tilde 0}$ and its orthogonal complement $\super Q
\equiv 1-\super P$. With help of $\super Q$ the well-defined
superoperator $\super R\equiv \super{QL}^{-1}\super Q$ represents
the pseudoinverse of the Liouvillean (``inverse on the regular
subspace $\super Q$").

With that we have all necessary ingredients for expressing the
current noise. It can be done in two different ways for the two
types of junctions. The $\R\L$-junction lies in the system and,
thus, the quantum regression theorem (see Ref.~\cite{gardiner},
Sec.~5.2) can be used to calculate the correlation
function~$C_{\R\L}(\tau)$. The final formula for the zero-frequency
current noise reads \cite{novotny}
\begin{equation}
    S_{\R\L}(0)
    = -2e^2\sbra{\tilde0}\super I_{\R\L}\super R\super
    I_{\R\L}\sket0,
    \label{noise_RL}
\end{equation}
where we have introduced the current superoperator
\begin{equation}
    \super I_{\R\L}\rho \equiv \frac1{2e}\{I_{\R\L},\rho\}
    \label{supercurrent_RL}
\end{equation}
in terms of which the stationary current is given as $\langle
I_{\R\L}\rangle = \tr\_S
\big(I_{\R\L}\rho\_{stat}\big)=e\sbra{\tilde0}\super
I_{\R\L}\sket0$.

For the outer junctions (between the dots and the leads) the QRT
cannot be used, because the current operators $I_{\L0}$ and
$I_{0\R}$ involve the lead operators. However, $n$-resolved form of
the generalized master equation and the MacDonald formula
\cite{robinson,elattari} enables us to calculate the zero-frequency
noise also for these junctions, see details in Ref.~\cite{novotny},
Sec.~III. For the stationary mean current through the $0\R$-junction
we get $\langle I_{0\R}\rangle = e\tr\_S\big(\super
I_{0\R}\rho\_{stat}\big) = e\sbra{\tilde0}\super I_{0\R}\sket0$ with
\begin{equation}
    \super I_{0\R}\rho = \Gamma\ket0\bra\R\rho\ket\R\bra0.
    \label{supercurrent_L0}
\end{equation}
The final result for the zero-frequency current noise reads
\begin{equation}
    S_{0\R}(0) = e^2\sbra{\tilde0}\super I_{0\R}-2\super I_{0\R}\super R\super I_{0\R}\sket0
    \label{noise_0R}
\end{equation}
and analogously for the $\L 0$-junction with $ \super I_{\L0}\rho =
\Gamma\ket\L\bra0\rho\ket0\bra\L$.

\section{Charge conservation issue}
\label{charge_conservation}

The equations of motion~\eqref{eom1_dqd} and~\eqref{eom2_dqd} for
the dot occupation operators (charge conservation conditions) imply
that the stationary mean current and the zero frequency noise are
independent of the measurement position along the
circuit~\cite{novotny}
$$
    \langle I_{\L0}\rangle = \langle I_{\R\L}\rangle =\langle I_{0\R}\rangle\,,\quad
    S_{\L0}(0) = S_{\R\L}(0) = S_{0\R}(0).
$$

Let us now focus on the reformulation of the charge conservation
condition in the superoperator language and evaluate, e.g., the
commutator $[\super N_\L, \super L]$ with the superoperator of
occupation of the left dot defined by $\super N_\L\rho =
\frac12\{n_\L,\rho\}$ in analogy with other superoperators
corresponding to system operators such as, e.g., $\super I_{\R\L}$.
Its matrix representation in the relevant 5-dimensional Liouville
subspace reads $\super N_\L=\mathrm{diag}(0,1,0,1/2,1/2)$. Then with
the help of Eqs.~\eqref{liouvillian_S_LR}, \eqref{LiouvillianB_LR},
and \eqref{LiouvillianL_LR} we arrive at
\begin{equation}
    [\super N_\L, \super L]=-\super I_{\R\L}-\super I\_{A}+\super I_{\L0},
    \label{super_charge_conservation0}
\end{equation}
with the ``anomalous current superoperator" $\super I\_{A}=-[\super
N_\L, \super L\_B]$. Since $[n_\L, H\_{BS}] =0$, we would expect
that also $[\super N_\L, \super L\_B]$ is zero, therefore the final
Liouville space analogy to the charge conservation
condition~\eqref{eom1_dqd} should read $[\super N_\L, \super L] =
\super I_{\L0}-\super I_{\R\L}$. Indeed, if this condition were
satisfied one could easily explicitly show the equivalence of
Eqs.~\eqref{noise_RL} and \eqref{noise_0R} using the method of
Ref.~\cite{novotny}, Sec.~IIID. Unfortunately, as one can readily
evaluate
\begin{equation}
    \super I\_{A} = \frac12\begin{pmatrix}
    0 & 0 & 0 & 0 & 0 \\
    0 & 0 & 0 & 0 & 0 \\
    0 & 0 & 0 & 0 & 0 \\
    0 & \gamma_+& \gamma_-& 0 & 0 \\
    0 & \gamma_+& \gamma_-& 0 & 0 \\
    \end{pmatrix},
    \label{anomalous_current}
\end{equation}
so that the anomalous current superoperator is not identically zero.
Analogously, for the other dot we obtain the result
\begin{equation}
    [\super N_\R, \super L] = \super I_{\R\L}+\super I\_{A}-\super
    I_{0\R},
    \label{super_charge_conservation2}
\end{equation}
with the opposite contribution form the heat bath compared to the
left dot, i.e., $[\super N_\R, \super L\_B] =-[\super N_\L, \super
L\_B] = \super I\_A$.

Thus, we face the fact that the charge conservation condition
contains nonzero anomalous terms. The possible consequences may be
that the mean current or the zero-frequency current noise are no
longer equal for all pairs of junctions. First, since
$\sbra{\tilde0}$ corresponds to the unity operator and, thus, is
equal to $(1, 1, 1, 0, 0)$, we see that $\sbra{\tilde0}\super
I\_A\sket0=0$ and the mean current is conserved along the whole
circuit. Also, we notice that $[\super N_\L+\super N_\R, \super
L\_B] = 0$ implies there is no problem with the charge conservation
between the outer junctions. However, the zero-frequency noise
eventually does show discrepancy between the outer junctions and the
inner one, in particular
\begin{align}
    &S_{\R\L}(0)-S_{\L0}(0) = e^2\sbra{\tilde0}\super I_{\L0}\super R\super
    I\_A\sket0=e^2\Gamma \Omega\epsilon\nonumber\\
    &\times\frac{\hbar^2\Gamma\gamma_+\Lambda^2+2\Omega^2\Lambda(\gamma_++\gamma_-)-2\epsilon^2\gamma_+(\frac{2\Omega}{\epsilon}\gamma_--\frac12\Gamma)}
    {\left[\frac12\Gamma\epsilon^2+3\Omega^2\Lambda+\frac12\hbar^2\Gamma\Lambda^2-\Omega\epsilon(2\gamma_++\gamma_-)\right]^2},
    \label{difference_S0}
\end{align}
where $\Lambda = \gamma_p+\frac12\Gamma$. We will analyze features
of this discrepancy in detail in the next section.

Before that, let us return to the point where we have assumed that
$[\super N_\L, \super L\_B]$ was identically zero. This was
a~reasonable assumption, because the dot occupation operator~$n_\L$
commutes with the heat bath-system interaction Hamiltonian~$H\_{BS}$
and thus the bath variables do not enter explicitly the current
operators. It can be shown that $[\super N_\L, \super L]$ indeed
does not depend on the bath variables if we operate on the whole
Liouville space (system + bath) before projecting onto the system
and introducing the weak coupling and Markovian
limits~\cite{diploma}. However, after the projection we arrive at
\begin{equation}
    \begin{aligned}
    &[\super N_\L, \super L\_B]\rho =\\
    &=\frac1{2\hbar^2}\int_0^\infty\d\tau\tr\_B\{\{[H\_{BS},n_\L],[H\_{BS}(-\tau),
    \rho\otimes\rho\_B]\}\}\\&+
    \frac1{2\hbar^2}\int_0^\infty\d\tau\tr\_B\{[H\_{BS},\{[H\_{BS}(-\tau),n_\L],
    \rho\otimes\rho\_B\}]\}.
    \end{aligned}
    \label{supercomm_NL_LB_int}
\end{equation}
Apparently, the first term is equal to zero due to $[n_\L, H\_{BS}]
=0$ as expected, while the second term yields the nonzero anomalous
current.

\section{Results}
\label{results}

In this section we present various aspects of the previously found
charge conservation breaking in the dissipative DQD model. In the
following we will conventionally represent the noise by
a~dimensionless quantity, the Fano factor $F = S(0)/e\langle
I\rangle$ \cite{blanter}.

\subsection{Zero dissipation}

In the case without the heat bath ($\gamma = 0$), the charge
conservation condition is satisfied. For the mean
current~\cite{nazarov} and the Fano factor~\cite{elattari} we obtain
\begin{gather*}
    \langle I\rangle = e\Gamma\,\frac{\Omega^2}{\epsilon^2+3\Omega^2+(\frac12\hbar\Gamma)^2}\,,\\
    F_{\L0} = F_{\R\L} = F_{0\R} = 1-\frac{4\Omega^2\(\Omega^2+2(\frac12\hbar\Gamma)^2\)}{\(\epsilon^2+3\Omega^2+(\frac12\hbar\Gamma)^2\)^2}\,.
\end{gather*}
These results are illustrated in Fig.~\ref{graf2}. The mean current
vs. bias~$\epsilon$ has the Lorentzian shape with the half-width
$\sqrt{3\Omega^2+(\hbar\Gamma/2)^2}$ and maximum at~$\epsilon =0$.
The Fano factor has the dip at~$\epsilon=0$ where quantum coherence
strongly suppresses the noise. For large $|\epsilon|>0$ the mean
current becomes very small and thus electrons tunnel very sparsely
and consequently the tunneling events are uncorrelated which
corresponds to a~Poisson process with the value of the Fano factor
$F\to1$.

\subsection{General case}

When dissipative heat bath comes into play ($\gamma>0$), the
transport is strongly affected by the possibility of exchanging
energy with the heat bath~\cite{aguado} as it is illustrated in
Fig.~\ref{graf2}. The shape of the mean current curve is no longer
Lorentzian but exhibits an asymmetry. With increasing temperature
the peak becomes broader and more symmetric. Analytically, we obtain
\begin{equation}
    \langle I\rangle =
    \frac{e\Gamma\Omega\(\Omega \Lambda-\gamma_+\epsilon\)}
    {\frac12\Gamma\epsilon^2+3\Omega^2\Lambda+\frac12\hbar^2\Gamma \Lambda^2-\Omega\epsilon(2\gamma_++\gamma_-)}.
    \label{mean_current_dissipation}
\end{equation}
In the right plot of Fig.~\ref{graf2} Fano factors $F_{\L0}=F_{0\R}$
(solid lines) and $F_{\R\L}$ (dotted lines) are plotted. The
difference between the Fano factors obtained by different approaches
is significant. Interestingly, at~$\epsilon=0$ the Fano factors have
the same value $F_{\L0}=F_{\R\L}$ as follows from
relation~\eqref{difference_S0}.

\obrazek[MP(),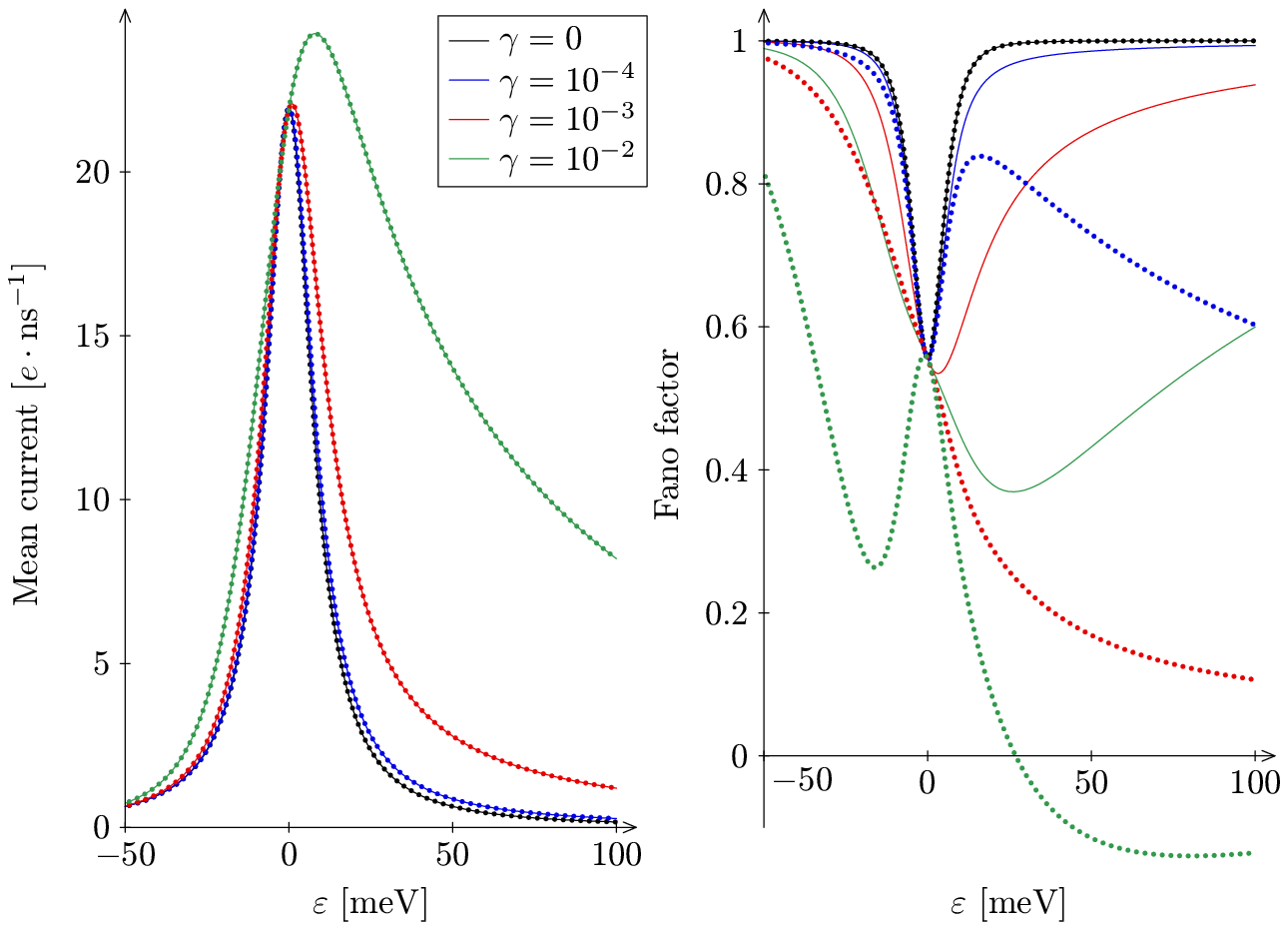,graf2]{Mean current and Fano
factor vs.~device bias~$\epsilon$ for different values of damping
coefficient~$\gamma$. Solid lines show the mean current and the Fano
factor at the outer junctions~$\L0$, $0\R$, dotted lines at the
inner junction~$\R\L$. Parameters: $\Omega = "5 meV"$, $\Gamma =
"0.1/\hbar{} meV"$, $\beta = "0.1 meV^{-1}"$.}

For $\epsilon>0$ spontaneous emission occurs even at very low
temperatures and the noise is reduced well bellow the Poisson limit.
Larger couplings~$\gamma$ lead to very asymmetric Fano factor. At
finite temperatures, absorption of energy quanta from the bath is
also possible and the Fano factor for $\epsilon<0$ is reduced bellow
the Poisson limit too. With increasing temperature the effect of the
emission and the absorption is growing, except the point
$\epsilon=0$ where both the mean current and the Fano factor are
temperature independent. It appears that the MacDonald formula
yields physically plausible results for~$F_{\L0}$ and~$F_{0\R}$,
whereas $F_{\R\L}$ given by the quantum regression theorem behaves
pathologically with unphysical negative values and non-Poisson limit
for $\epsilon\to\infty$. For sufficiently strong coupling
($\gamma\approx10^{-2}$) $F_{\R\L}$ drops to negative values in the
$\epsilon>0$ region and for sufficiently high temperature
($T\approx"200 K"$) also in the $\epsilon<0$ region. Analyzing the
expression~\eqref{difference_S0} we find that for
$\epsilon\to\infty$ the noise difference $\Delta S =
S_{\R\L}(0)-S_{\L0}(0)\sim 1/\epsilon$. From the
relation~\eqref{mean_current_dissipation} for the mean current in
the same limit it follows $\langle I\rangle\sim 1/\epsilon$.
Therefore their ratio yielding the difference of Fano factors
$\Delta F = F_{\R\L}-F_{\L0}$ does not go to zero for
$\epsilon\to\infty$ as it should. Nevertheless, despite of the fact
that $F_{\R\L}$ behaves manifestly wrong, we do not have any valid
proof yet, that the MacDonald formula gives a~better and more
reliable results for~$F\_{L0}$ and $F_{0R}$.

In the following subsections we will investigate how both the
MacDonald formula and the quantum regression theorem approach behave
in several approximations or limit cases. It will answer whether the
MacDonald formula gives physically acceptable results and will show
more pathologies of the quantum regression theorem results.

\subsection{Limit ${\Gamma\to0}$}

Limit $\Gamma\to0$ could be potentially interesting~-- analogously
with the dissipationless limit $\gamma\to0$ the charge conservation
could possibly be recovered. The mean current and the zero-frequency
noise go to zero in this limit, however their ratio, the Fano
factor, does not~\cite{diploma}. The difference between the Fano
factors of the outer and inner junctions reads
\begin{multline*}
    F\_{L0}-F\_{RL} =\hbox{}\\=
    \frac{2\epsilon^2\(\cotgh^2{(\frac12\beta\Delta)}-1\)}
    {\epsilon^2+4\epsilon\Delta\cotgh{(\frac12\beta\Delta)}+3\Delta^2\cotgh^2{(\frac12\beta\Delta)}}.
\end{multline*}
Results are illustrated in Fig.~\ref{graf3}. We note quite an
interesting phenomenon that the Fano factor does not depend on the
heat bath spectral density and, thus, it is not influenced by the
strength~$\gamma$ of the dissipation. Nevertheless, all the
anomalies survive. The Fano factor~$F_{\R\L}$ can be negative for
certain~$\epsilon$ and temperature high enough. Since the Fano
factors for the three junctions are not equal, the charge
conservation condition is not fulfilled. Both $F_{\L0}$
and~$F_{\R\L}$ become~1 for $\epsilon\to-\infty$, $1/2$ for
$\epsilon\to\infty$ and $F = 5/9$ for $\epsilon=0$. The striking
difference between the Fano factors for the inner and outer
junctions is that $F_{0\L}$ has no maxima or minima and just
smoothly decreases from 1~to~1/2, whereas $F_{\R\L}$ has two minima,
one for $\epsilon<0$ and the other for $\epsilon>0$.

\obrazek[MP(),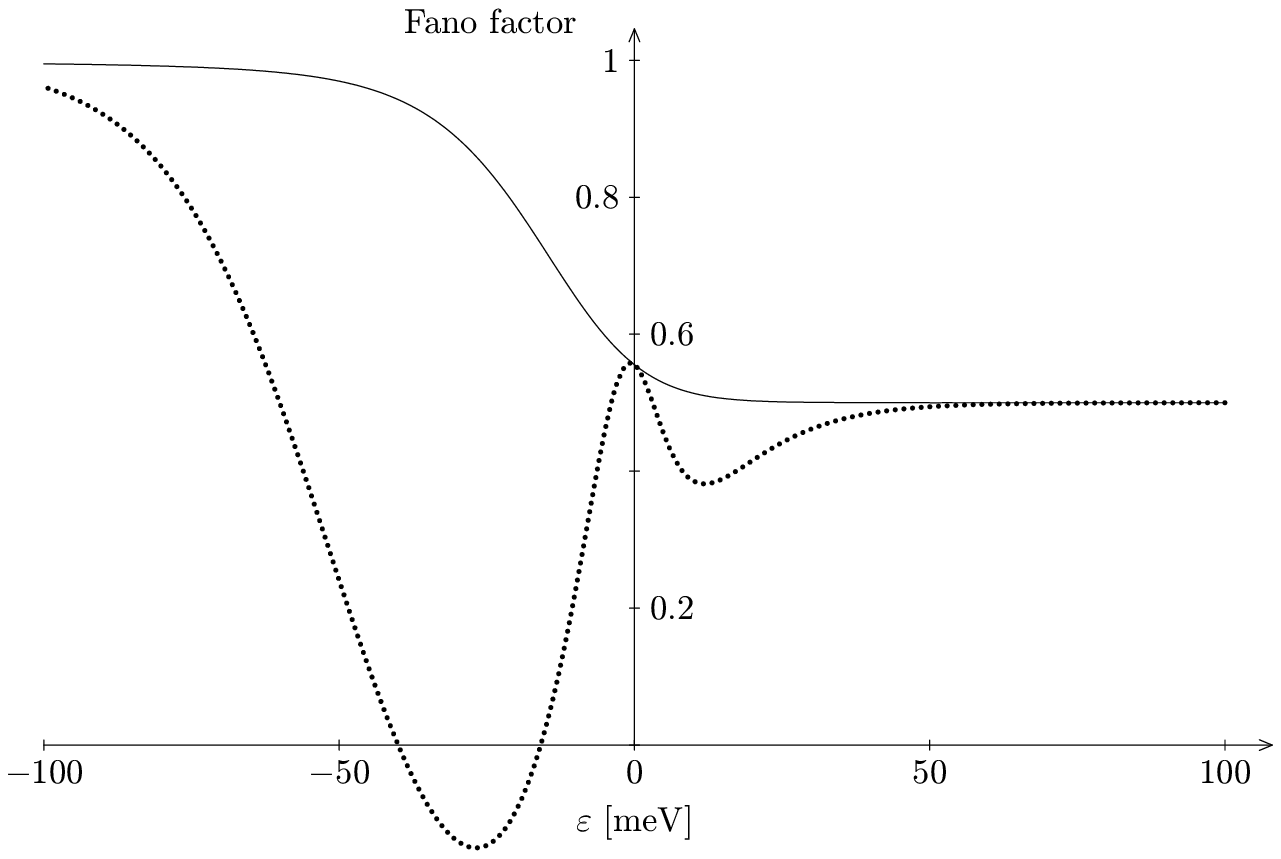,graf3]{Fano factor vs.~device
bias~$\epsilon$ for $\Gamma\to0$. Solid line shows
$F_{\L0}=F_{0\R}$, dotted line $F_{\R\L}$. Parameters: $\Omega = "5
meV"$, $\beta = "0.1 meV^{-1}"$.}

\subsection{Rotating wave approximation}
\label{dqd:section:rwa}

The rotating wave approximation (RWA) is understood as neglecting
terms of the form $\sigma^\dagger \rho \sigma^\dagger$,
$\sigma\rho\sigma$ and keeping terms of the form $\sigma^\dagger\rho
\sigma$, $\sigma\rho \sigma^\dagger$ in the generalized master
equation $\d\rho/\d t =\super L\rho$ \cite{gardiner}. In our
language, we mean $\sigma^\dagger \equiv \ket\L\bra\R$,
$\sigma\equiv \ket\R\bra\L$ and, therefore, $\ket\L\bra \L =
\sigma^\dagger\sigma$, $\ket\R\bra\R = \sigma\sigma^\dagger$.
Application of the RWA to~\eqref{LiouvillianB_LR} leaves us with
\begin{equation*}
    \super L\_B^{\rm RWA} = \begin{pmatrix}
    0 & 0 & 0 & 0 & 0 \\
    0 & 0 & 0 & 0 & 0 \\
    0 & 0 & 0 & 0 & 0 \\
    0 & 0 & 0 & -\gamma_p& 0 \\
    0 & 0 & 0 & 0 & -\gamma_p\\
    \end{pmatrix}.
\end{equation*}
The anomalous current~\eqref{anomalous_current} is now identically
zero, since $[\super N_\L, \super L\_B^{\rm RWA}] = [\super
N_\R,\super L\_B^{\rm RWA}] = 0$, and therefore the charge
conservation is restored. The zero-frequency noise and the Fano
factor for all junctions are equal. The mean current is
$$
    \langle I\rangle =
    \frac{e\Gamma\Omega^2\Lambda}{\frac12\Gamma\epsilon^2+3\Omega^2\Lambda+\frac12\hbar^2\Gamma \Lambda^2}
$$
and for the Fano factor we get
$$
    F =
    1-\frac{4\Omega^2\left[\Lambda^2\(\Omega^2+\frac12\hbar^2\Gamma(\gamma_p+\Gamma)\)+\frac12\gamma_p\Gamma\epsilon^2\right]}
    {\left[\frac12\Gamma\epsilon^2+3\Omega^2\Lambda+\frac12\hbar^2\Gamma\Lambda^2\right]^2}\,.
$$
These results are illustrated in Fig.~\ref{graf5}.

\obrazek[MP(),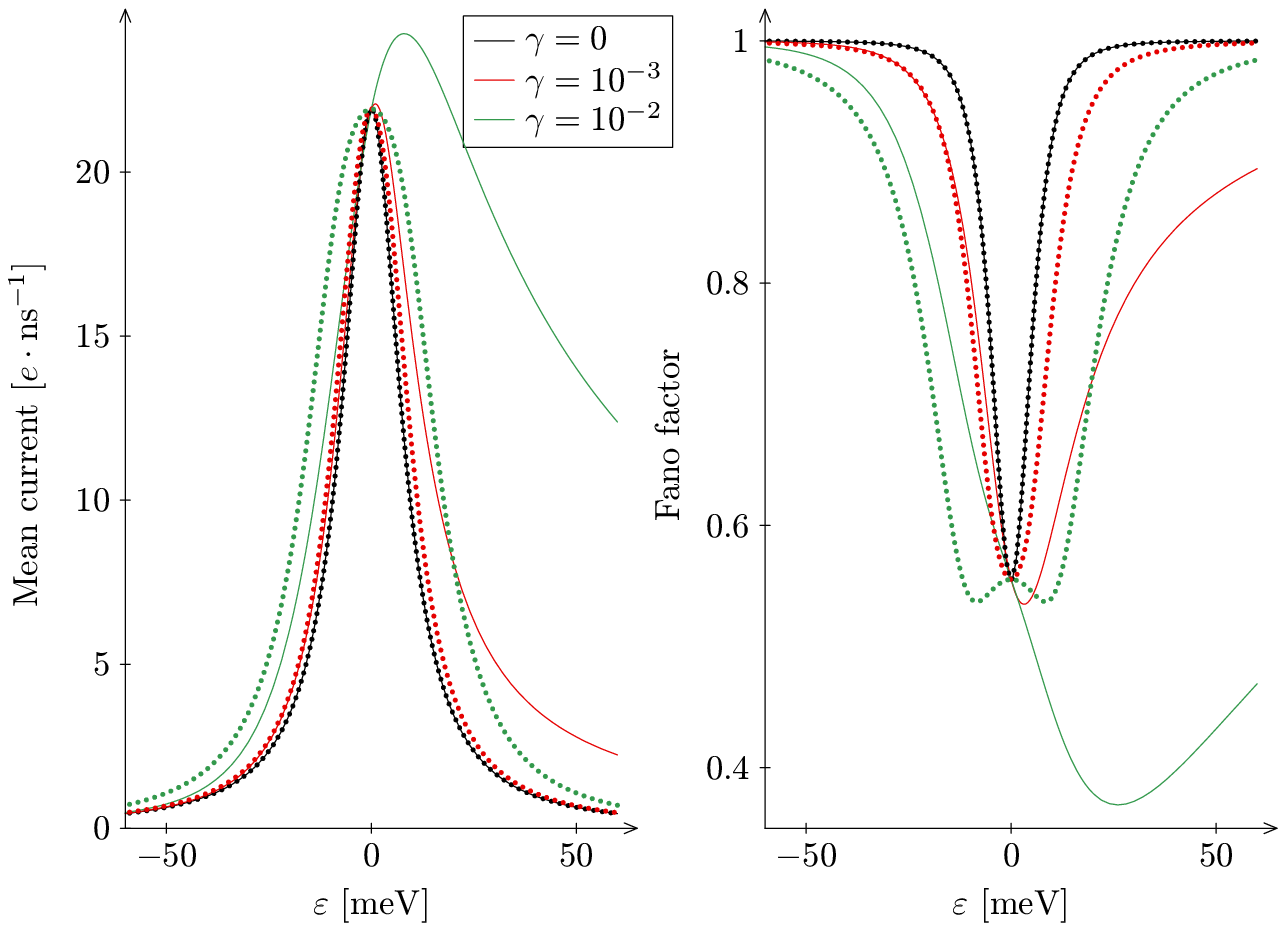,graf5]{Mean current and Fano factor
vs.~device bias~$\epsilon$ for different values of damping
coefficient~$\gamma$. Solid lines correspond to the outer junctions
without the RWA while dotted lines correspond to all junctions
within RWA. Parameters: $\Gamma = "0.1/\hbar{} meV"$, $\Omega = "5
meV"$, $\beta = "0.1 meV^{-1}"$.}

We can see that negative Fano factor does not appear in this
approach. Furthermore, all junctions give identical results for both
the mean current as well as the current noise. Thus, it seems that
all problems are fixed. However, the physical content of our results
has undergone great changes. We have given reasons why the mean
current and the Fano factor curves should have emission-absorption
asymmetry, but we obtain absolutely symmetric curves~-- the mean
current gained the Lorentzian shape and the Fano factor has no
suppression for $\epsilon>0$ due to the emission process. Because of
these reasons we must reject the rotating wave approximation on the
physical grounds.

\subsection{Pauli master equation}
\label{dqd:section:pauli}

Pauli master equation is yet another modification of the original
GME which does not lead to formal inconsistencies. When the coupling
to the leads $\Gamma$ is small enough, which is the case in the
regime we consider here, we can neglect the off-diagonal elements
$\rho_{12},\,\rho_{21}$ of the density matrix in the system
eigenbasis \cite{spohn}. If we transform the Liouvillian of the
original GME in the left-right basis \eqref{dqd:gme} into the system
eigenbasis $\{\ket 0,\,\ket 1,\,\ket 2\}$, we can get the
Liouvillian (or rate matrix) of the Pauli master equation
immediately just by restricting ourselves to the subspace spanned by
$\rho_{00},\,\rho_{11},\,\rho_{22}$ which leads to
$$
    \super L^{\rm Pauli} = \begin{pmatrix}
    -\Gamma_{10}-\Gamma_{20} & \Gamma_{01} & \Gamma_{02} \\
    \Gamma_{10} & -\Gamma_{01}-\gamma_\downarrow & \gamma_\uparrow  \\
    \Gamma_{20} & \gamma_\downarrow & -\Gamma_{02}-\gamma_\uparrow  \\
    \end{pmatrix},
$$
where
\begin{gather*}
    \Gamma_{10} = \Gamma_{02} = \Gamma\,\frac{\Delta+\epsilon}{2\Delta},\qquad
    \Gamma_{01} = \Gamma_{20} =
    \Gamma\,\frac{\Delta-\epsilon}{2\Delta},\\
    \gamma_\downarrow = \frac{4\pi}{\hbar^2}\frac{\Omega^2}{\Delta^2}\,J(\Delta/\hbar)\,\frac1{1-\eu^{-\beta\Delta}},\\
    \gamma_\uparrow = \frac{4\pi}{\hbar^2}\frac{\Omega^2}{\Delta^2}\,J(\Delta/\hbar)\,\frac1{\eu^{\beta\Delta}-1}.
\end{gather*}
Identical results can be derived directly from a rate equation
approach to the occupations of the eigenstates only with the rates
determined by the Fermi golden rule. The current superoperators in
the new Liouville subspace are given by
$$
    \super I_{\L0} = \begin{pmatrix}
    0 & 0 & 0 \\
    \Gamma_{10} & 0 & 0 \\
    \Gamma_{20} & 0 & 0 \\
    \end{pmatrix},\qquad
    \super I_{0\R} =\begin{pmatrix}
    0 & \Gamma_{01} & \Gamma_{02} \\
    0 & 0 & 0 \\
    0 & 0 & 0 \\
    \end{pmatrix},
$$
while no analogy of the superoperator~$\super I_{\R\L}$ exists on
the chosen subspace.

For the mean current and the Fano factor between the dot and the
lead, we get regardless of the choice of the left or right junction
$$
    \langle I\rangle = \frac{e\Gamma\Omega\(\Omega\Lambda-\gamma_+\epsilon\)}
    {\frac12\Gamma\epsilon^2+3\Omega^2\Lambda-\Omega\epsilon(2\gamma_++\gamma_-)}
$$
and
\begin{multline*}
    F =1-2\Omega\times\hbox{}\\
    \times\frac{2\Omega^3\Lambda^2-\Omega^2\epsilon(\gamma_p+\frac32\Gamma)(3\gamma_++\gamma_-)-\gamma_+\epsilon^3\Lambda}
    {\left[\frac12\Gamma\epsilon^2+3\Omega^2\Lambda-\Omega\epsilon(2\gamma_++\gamma_-)\right]^2}.
\end{multline*}
These expressions differ from the
result~\eqref{mean_current_dissipation} and the corresponding
expression for $F_{L0},\,F_{0R}$ for the full-space reduced density
matrix in the second or the first order of~$\Gamma$, respectively,
i.e., they both agree in the lowest order in $\Gamma$ as should be
expected. Thus, for small values of~$\Gamma$ we obtain very good
agreement between the two approaches (curves are almost
indistinguishable from the solid lines in Fig.~\ref{graf2}). This
finding finally justifies the results obtained by the full approach
as physically plausible although it does not yield any hints where
the problem of the full approach might lie nor does it say anything
reliable for larger $\Gamma$'s.


\section{Discussion and conclusions}
\label{conclusions}

We have presented our results for the dissipative DQD system and
explicitly pointed out the paradoxes stemming from the Markovian GME
description of this system, such as charge conservation breaking
between different junctions or unphysical negative values of the
current noise. The weak coupling prescription is known to possess
severe conceptual problems \cite{gardiner,kohen} including the
ambiguity of the choice of the kernel (e.g., direct weak coupling
prescription \`{a} la Bloch-Redfield vs.\ RWA/secular approximation
on top of that), breaking of positivity of the reduced density
matrix within the Bloch-Redfield formalism, and breaking of the
equations of motion (Ehrenfest theorem) within the RWA
\cite{oconnell}. Thus, in some sense one shouldn't be surprised to
find similar inconsistencies in the DQD study. Yet, there are also
important differences between the above mentioned effects and the
present findings such as relative importance of the various
discrepancies even for very small coupling constants, i.e.,
significantly increased sensitivity of these effects with the
presence of the other bath. This indicates that the phenomena
encountered here may be going beyond the ``standard" weak coupling
paradoxes and are specific to multiple-bath setups.

This conjecture seems to be supported by further facts: the charge
conservation is fulfilled without the presence of the dissipative
bath which is a necessary condition for the occurrence of the
breaking. RWA applied to our system expectedly breaks the equations
of motion (physically incorrect symmetry between the emission and
absorption processes), however, the full Bloch-Redfield kernel not
only disobeys the positivity (negative noise) but also breaks the
equations of motion (charge conservation breaking) which does not
seem to follow the standard weak coupling behavior. The
multiple-bath dissipative systems have been known as challenging for
quite long but at the same time they are characterized by physically
interesting and sometimes counterintuitive behavior
\cite{burkard,kohler} which can be also responsible for the present
findings. Moreover, as a part of the diploma thesis of the first
author \cite{diploma} other systems were studied, e.g., the energy
transport in two linearly coupled harmonic oscillators being an
exactly solvable analog to the dissipative DQD but no similar
phenomenon was observed there. This leads to speculations that these
findings are determined not only by the simultaneous presence of
more baths but also by the non-Gaussian character of the associated
noise missing in the linear systems.

To sum up, although the reported issue resembles the notorious
difficulties and paradoxes of the weak coupling theory, neither its
exact origin nor possible cures have been uniquely identified so
far. Yet, we present our findings to the community in order to draw
attention to this open nontrivial problem which may be, apart from
its relevance for the particular physical system under study, of
interest in the broader context of open dissipative quantum systems.
In particular, our findings raise the questions of the development
of charge-conserving approximation schemes within the generalized
master equation approaches analogous to their non-equilibrium
Green's function counterparts and of general understanding of
dynamics of quantum systems coupled to multiple baths.

\section*{Acknowledgments}

The authors thank K.~Neto\v cn\'y and B.~Velick\'y for stimulating
discussions and comments. This work was supported by Grant No.~62108
of the Grant agency of Charles University in Prague which the
authors gratefully acknowledge. The work of T.~N.\ is a part of the
research plan MSM 0021620834 financed by the Ministry of Education
of the Czech Republic.

\def\refname{References}

\end{document}